\documentclass[aps,prd,preprint,groupedaddress]{revtex4}
\usepackage{latexsym}
\usepackage{graphics}

\usepackage{graphicx}
\usepackage{dcolumn}
\usepackage{bm}
\usepackage{amsmath}
\usepackage{slashbox}

\def\be{\begin{equation}}
\def\ee{\end{equation}}

\begin{document}
\title{Confronting fragmentation function universality with single hadron inclusive production at
HERA and $e^+ e^-$ colliders}
\author{S.~Albino}
\author{B.~A.~Kniehl}
\author{G.~Kramer}
\author{C.~Sandoval}
\affiliation{{II.} Institute for Theoretical Physics, University of Hamburg,\\
             Luruper Chaussee 149, 22761 Hamburg, Germany}
\date{November 2, 2006}
\begin{abstract}
Predictions for light charged hadron production data in the current fragmentation region of
deeply inelastic scattering from the H1 and ZEUS experiments 
are calculated using perturbative Quantum Chromodynamics at next-to-leading order,
and using fragmentation functions obtained by fitting to similar data from
$e^+ e^-$ reactions. General good agreement
is found when the magnitude $Q^2$ of the hard photon's virtuality is sufficiently
large. The discrepancy at low $Q$ and small scaled momentum $x_p$ is reduced by incorporating
mass effects of the detected hadron. By performing quark tagging, 
the contributions to the overall fragmentation
from the various quark flavours in the $ep$ reactions are studied and compared
to the contributions in $e^+ e^-$ reactions. The yields of the various hadron species
are also calculated.
\end{abstract}
\maketitle

\section{Introduction}
\label{intro}

Due to their high accuracy, data for single hadron inclusive production
in high energy $e^+ e^-$ reactions have been used within the framework
of the factorization theorem of Quantum Chromodynamics (QCD) at
leading twist and at next-to-leading order (NLO) to constrain
fragmentation functions (FFs) for charge summed light charged hadrons
($\pi^\pm$, $K^\pm$ and $p/\bar{p}$) in Refs.\
\cite{Kniehl:2000fe,Kretzer:2000yf,Bourhis:2000gs,Albino:2005me}.
The benefits of such an extraction are twofold. First, a test of
perturbative QCD is provided and consequently imposes a constraint on
the strong coupling constant $\alpha_s (M_Z)$ at the $Z$ boson mass scale
$M_Z$. Second, since the universality principle of the
factorization theorem implies that the FFs are independent of the
initial state, FFs extracted in this way can be used to make
predictions for other hadron production processes such as those
arising from $ep$ reactions in the current fragmentation region and
from $p p$ and $p\bar{p}$ reactions.

Tests of universality were performed in Ref.\ \cite{Kniehl:2000hk} by
confronting predictions obtained from the KKP FF set
\cite{Kniehl:2000fe} with corresponding measurements of rapidity ($y$) and
transverse momentum ($p_T$) distributions for unidentified light charged
hadron production in $p\bar{p}$ reactions at UA1, UA2 and CDF, $\gamma
p$ reactions at H1 and ZEUS, and $\gamma \gamma$ reactions at
OPAL. Within the theoretical and experimental errors the description
of all data sets was good. However, the predictions for the $p\bar{p}$
reactions have large theoretical errors arising from scale variations,
and the experimental errors are largest at large $p_T$ where additional
non-perturbative information such as higher twist is expected to be least important. The $\gamma p$
and $\gamma \gamma$ reaction data suffer from similar problems, but in
addition the predictions gain large errors from the rather badly
constrained parton distribution functions (PDFs) of the photon. 
Distributions in $y$ generally have even larger theoretical errors. More
stringent tests of universality were performed in Refs.\
\cite{Kniehl:2004hf,Daleo:2004pn,Aurenche:2003by} through analysis of
pseudorapidity and $p_T$ distributions from H1 for the
process $ep \rightarrow e+\pi^0 +X$, and in Ref.\ \cite{Kniehl:2004hf}
through analysis of $p_T$ distributions from ZEUS for
the process $ep \rightarrow e+h +X$, which did not require the use of
photon PDFs (except in the low $Q$ region
\cite{Fontannaz:2004ev}). The disagreement found with the ZEUS data was
reduced in Ref.\ \cite{Nadolsky:2000ky} through resummation of
multiple parton radiation at low $p_T$.

In this paper we confront predictions of normalized light charged
hadron scaled momentum ($x_p$) distributions with single hadron inclusive production measurements
in deeply inelastic scattering at the H1
\cite{Adloff:1997fr} and ZEUS \cite{Derrick:1995xg,Breitweg:1997ra} experiments
(the more recent ZEUS data of Ref.\ \cite{Breitweg:1999nt} are
unfortunately unavailable) at high $Q$ in the current
fragmentation region, where the detected hadron originates from the
fragmentation of a parton at high scale. These hadrons can be reliably
distinguished from those in the target fragmentation region by working
in the Breit frame, where the struck quark, which subsequently fragments, moves in the opposite
direction to the proton remnants, so that $x_p$
distributions in the current fragmentation region are closely related
to $x_p$ distributions in any one of the two event hemispheres of
$e^+ e^-$ reactions. Consequently, comparison of predictions for $ep$ reaction data 
using FFs constrained from $e^+ e^-$ reaction data allows for more
direct tests of universality. Since the data are normalized,
uncertainties from the proton PDFs and their perturbative evolution
are reduced, as well as the dependence on Bjorken $x$.

The charge-squared weighted FFs are weighted equally in $e^+ e^-$
reactions. In particular, this implies that FFs for massless $d$ and $s$ quarks cannot be
separated if they are not separately tagged, so that, since no
individually tagged light quark flavour data was used in the analyses
of Refs.\ \cite {Kniehl:2000fe,Kretzer:2000yf,Bourhis:2000gs}, additional theoretical
constraints on the $d$ quark had to be imposed. However, calculation of hadron
production processes from proton initiated reactions at facilities
such as HERA ($e^+ p$), the Tevatron ($p\bar{p}$), RHIC and the LHC ($pp$), where the
charge-squared weighted FFs for quarks of each flavour have an
independent weighting provided by the PDFs, may demand some degree of
knowledge of the individual quark FFs, particularly in the light quark
sector.

In the determination of the AKK FF set for light charged hadrons 
\cite{Albino:2005me}, a more phenomenological separation of
the light quark flavour FFs was pursued using the individually quark
flavour tagged probabilities measured by the OPAL collaboration
\cite{Abbiendi:1999ry}. These probabilities were constrained by
single and double hadron inclusive production measurements for
which light quarks are favoured, together with the well justified
theoretical assumptions of SU(2) isospin invariance between $u$ and
$d$ quarks for the quark compositions of $\pi^\pm$, and the
branching ratios of the $Z$ boson into quark-antiquark pairs of each
quark flavour from perturbative QCD. Small $x_p$ subtleties in double
hadron inclusive production are relatively unimportant since the data
are in the range $x_p > 0.2$. Such a separation should make little
difference to the current knowledge of $\pi^\pm$ FFs, since the SU(2)
isospin relation was also used in the extraction of the KKP FF set to
constrain the $d$ quark. Thus, predictions for $\pi^\pm$
data should not depend too much on the choice of FF set. 
The same applies to unidentified light charged hadron data, albeit to a slightly less degree,
since $\pi^\pm$ dominates the sample on account of its low mass.
However, the anticipated strange quark
suppression in $K^\pm$ production observed in the OPAL experiment
gave more realistic $K^\pm$ production FFs for $d$ and $s$ quarks in the AKK set than those in
the KKP set, where the FF for $s$ was set
equal to the FF for $u$ for simplicity. Finally, the light quark separation of the AKK
FFs for $p/\bar{p}$ production may also be significant,
although it was limited by the large experimental uncertainties of the
OPAL tagging probabilities.

These expectations are found to some degree in the comparisons of
theoretical predictions with $pp$ initiated single hadron inclusive production data
\cite{Adams:2006nd,Abelev:2006cs} from the STAR collaboration. In
Ref.\ \cite{Adams:2006nd}, both the AKK and KKP
FF sets lead to similar and good descriptions of the $\pi^\pm$ yield,
while the theoretical prediction gives better agreement
(at scale $\mu=p_T$) with the measured $p/\bar{p}$
production when the $p/\bar{p}$ FFs are employed from the AKK set than
from the KKP set. The AKK set for $K^\pm$ and $K_S^0$ \cite{Albino:2005mv} also
resulted in an improvement \cite{Heinz:2005vu} in the theoretical
description of the $K_S^0$ production measurements of Ref.\
\cite{Abelev:2006cs}.

The paper is organized as follows. We first present the formalism
behind our calculations in section \ref{TF}. We define
the observable we are studying, and give the form of the cross section in terms of the FFs
to underline the similarities among, and differences between, single
hadron inclusive production in $e^+ e^-$ and $ep$ reactions. Then we discuss the
modification to the cross section when the detected hadron's mass
is not negligible, since this effect is important at sufficiently small $x_p$
and low $Q$. Section \ref{compdata} contains our comparisons
with the data, and we examine the uncertainties arising from the arbitrary
choice of scale, of PDF set and of FF set, as well as the importance of gluon
fragmentation and of the detected hadron mass effect. Furthermore, although
the corresponding data is absent, the 
contributions from the individual fragmenting parton and detected
hadron species to the cross section are calculated to further determine differences and
similarities of the FF sets. In section \ref{conc} we present
our conclusions. Finally, the appendix gives details on the cuts used
in the experiments.

\section{Theoretical formalism}
\label{TF}

\begin{figure}
\begin{center}
\includegraphics[width=8.5cm]{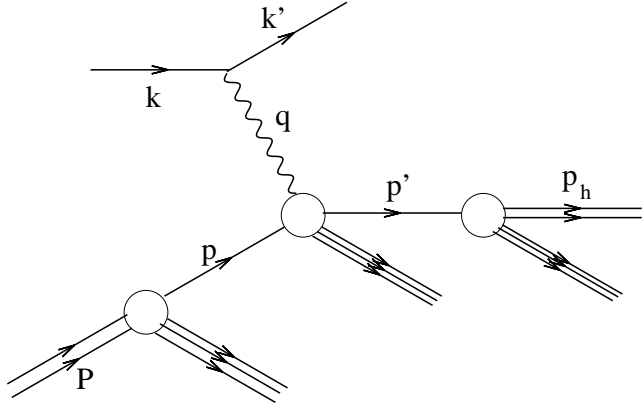}
\caption{General graph for the leading twist
contributions to the process $e(k)p(P)\rightarrow e(k')+h(p_h)+X$. Parallel trios of
lines signify unobserved final states.}
\label{kinematics}
\end{center}
\end{figure}

We are concerned with the process $ep\rightarrow e+h+X$, where $h$ is a detected hadron and
$X$ is the remaining unobserved part of the final state, whose kinematic variables will be assigned
according to the external particles of the general graph in Fig.\ \ref{kinematics}.
The kinematic degrees of freedom are chosen to be the centre-of-mass (c.m.) energy $\sqrt{s}$ of the
initial state electron-proton system,
which is given by $s=(P+k)^2$ and which is kept fixed in the experiments, 
the magnitude of the hard photon's virtuality
$Q^2=-q^2$, the Bjorken scaling variable $x=Q^2/(2P\cdot q)$ and the scaled
detected hadron momentum $x_p=2p_h \cdot q/q^2$. The 
normalized cross section (with the $s$ dependence omitted for brevity) takes the form
\be
F^{{\rm proton}\ h}({\rm cuts},x_{pA},x_{pB})=\frac{\int_{\rm cuts} dQ^2 dx \int_{x_{pA}}^{x_{pB}} dx_p 
\frac{d {\mathcal O}^{{\rm proton}\ h}}{dx_p}(x,x_p,Q^2)}
{\int_{\rm cuts} dQ^2 dx {\mathcal O}^{\rm proton} (x,Q^2)},
\label{XSmeas}
\ee
where, for convenience later, we use the shorthand ${\mathcal O}$ for $d^2 \sigma/(dx dQ^2)$,
where ``cuts'' refers to a specified region in the $(x,Q^2)$ plane (see the appendix for 
the various cuts used by H1 and ZEUS), and where $x_{pA(B)}$
is the lower (upper) edge of the $x_p$ bin.
The cross section and the kinematic variables
are frame invariant, and are measured in the Breit frame, defined to be
the frame where the photon energy vanishes.
In this frame the target fragmentation region ($x_p <0$) contains the proton remnants, while
the struck parton fragments into the current fragmentation region ($x_p > 0$), and the latter process
is equivalent to the fragmentation of a parton into an event hemisphere in $e^+ e^-$ reactions.

\subsection{QCD factorization theorem}

The factorization theorem dictates that the leading twist component of
the factorized cross section in the numerator of Eq.\ (\ref{XSmeas}) is calculated from processes
of the form shown in Fig.\ \ref{kinematics} and takes the form
\be
\begin{split}
\frac{d {\mathcal O}^{{\rm proton}\ h}}{dx_p}(x,x_p,Q^2)
=\int_x^1
\frac{dy}{y} \int_{x_p}^1 \frac{dz}{z} \sum_{ij} \frac{d \widehat{\mathcal O}^{ij}}{dz}
&\left(y,z,\frac{Q^2}{\mu^2},a_s(\mu^2)\right) \\
&\times \frac{x}{y} f_i^{\rm proton} \left(\frac{x}{y},\mu^2\right)\frac{x_p}{z} D_j^h \left(\frac{x_p}{z},\mu^2\right).
\label{xsinfacttheor}
\end{split}
\ee
In this framework, the incoming parton $i$ has momentum $p=(x/y)P$ and the outgoing
parton $j$ has momentum $p'=(z/x_p) p_h$.
$f_i^{\rm proton}$ is the PDF of parton $i$ in the proton,
$D_j^h$ is the FF of parton $j$ to the hadron $h$, $\widehat{\mathcal O}^{ij}$ is the equivalent
factorized partonic observable given to NLO in Ref.\ \cite{Altarelli:1979kv}, $\mu$ is the factorization / 
renormalization scale which distinguishes the soft from the hard subprocesses and
$a_s(\mu^2)=\alpha_s (\mu)/(2\pi)$. 
The more commonly written form of Eq.\ (\ref{xsinfacttheor}) can be obtained by changing
the integration variables to $\widehat{x}=x/y$ and $\widehat{x}_p= x_p/z$, which accentuates
the role of the PDFs and FFs as probability densities.
Using the momentum sum rule
\be
\sum_h \int_0^1 dz z D_j^h (z,\mu^2)=1,
\ee
the integration over $x_p$ from 0 to 1 and the sum over $h$ of Eq.\ (\ref{xsinfacttheor}) yields
the factorized cross section in the denominator of Eq.\ (\ref{XSmeas}), viz.\
\be
{\mathcal O}^{\rm proton}(x,Q^2)=\int_x^1
\frac{dy}{y} \sum_{i} \widehat{\mathcal O}^i
\left(y,\frac{Q^2}{\mu^2},a_s(\mu^2)\right) \frac{x}{y} f_i^{\rm proton} \left(\frac{x}{y},\mu^2\right).
\label{xsinfacttheordenom}
\ee

\subsection{Comparison with $e^+ e^- \rightarrow h+X$}
\label{compepem}

We now perform a pedagogical study of the FF dependence of $F^{{\rm
proton}\ h}$. Therefore, and in this subsection only, we work to leading  
order (LO) (however, all calculations used for our numerical analysis of section
III will be performed to NLO), where
\be
\frac{d \widehat{\mathcal O}^{ij}}{dz}
\left(y,z,\frac{Q^2}{\mu^2},a_s(\mu^2)\right)
=\frac{d\sigma_0}{dQ^2} (Q^2)\sum_I \delta_{ij} \delta_{iI}
e_{q_I}^2(Q^2) \delta(1-y) \delta(1-z)
\ee
and
\be
\widehat{\mathcal O}^i
\left(y,\frac{Q^2}{\mu^2},a_s(\mu^2)\right)=\frac{d\sigma_0}{dQ^2} (Q^2)
\sum_I \delta_{iI} e_{q_I}^2(Q^2) \delta(1-y).
\ee
In these expressions, $\sigma_0$ is the cross section for the elastic process
$e \mu \rightarrow e \mu$ for one photon exchange in the $t$-channel, and
$I$ indexes the quark of effective electroweak charge $e_{q_I}(Q^2)$.
As a further simplification, we will neglect the bin width in $Q$
since $F^{{\rm proton}\ h}$ is approximately independent of $Q$ up to $O(1/\ln Q)$ corrections. 
(However, we will not neglect the
bin widths in $Q$ in section \ref{compdata}.) The result is
\be
F^{{\rm proton}\ h}({\rm cuts},x_{pA},x_{pB})
=\frac{\int_{x_{pA}}^{x_{pB}} dx_p \sum_I e_{q_I}^2(Q^2) G_I(Q^2)
x_p D_I^h (x_p,Q^2)}{\sum_J e_{q_J}^2(Q^2) G_J(Q^2)},
\label{simpformofF}
\ee
where $G_I(Q^2)=\int_{\rm cuts} dx\ x f_I^{\rm proton}(x,Q^2)$.
If the $G_I$ are independent of $I$, the numerator of Eq.\ (\ref{simpformofF}) is equal to
the equivalent LO result for $e^+ e^- \rightarrow h+X$. 
It is therefore essentially the differences between the 
$G_I$ which distinguishes the two types of observables. 
The relative sizes of the components of the $ep$ cross section where the 
quark directly connected to the electroweak boson is tagged
help to determine the relative importance of the fragmentation of individual quark flavours 
in the untagged cross section in $ep$ reactions, and consequently to what extent 
these data could complement the untagged and tagged
data from $e^+ e^-$ reactions in understanding fragmentation from the various quark flavours.
The $ep$ cross section for which quark $I$ is tagged 
can be obtained by setting the remaining quark charges to zero
in the calculation, implying that it is scale independent and is given at LO
by $e_{q_I}^2(Q^2) G_I(Q^2) x_p D_I^h (x_p,Q^2)$. Then
the largest component is the tagged cross section for which $I=u$, due to the valence
structure of the proton, the larger charge of the $u$ quark relative to 
the $d$ quark and, to some extent, because
the $u$ quark is the most favoured one in the production of light charged hadrons. 
By the nature of data from $e^+ e^-$ reactions,
the $u$ quark fragmentation is currently also the most constrained component, 
in particular for $\pi^\pm$ production which constitutes most of the sample,
while the most unconstrained component is the difference between the 
$d$ and $s$ quark fragmentations due to their similar effective electroweak charges. 
Consequently, in the absence of sufficiently precise data for $ep$ reactions, 
$uds$ (or equivalently $c$ and $b$) tagging would therefore be valuable since 
together with $e^+ e^-$ reaction data it would provide some constraint on 
the separation between $d$ and $s$ quark fragmentation, by virtue of the difference between
$G_d$ and $G_s$. So far only charm quark tagging through $D^{*\pm}$ production measurements
\cite{Chekanov:2003rb} has been performed in the fully inclusive case
to obtain $F_2^c$. Separate $u$, $d$ and $s$ tagging as performed in $e^+ e^-$ 
reactions \cite{Abbiendi:1999ry} would constitute a further improvement,
but may not be possible at present.

Gluon fragmentation is not so well constrained by $e^+ e^-$ reactions since the gluon
does not couple directly to the electroweak boson. Although 
the proton is an abundant source of gluons,
this uncertainty is unlikely to contaminate the measurements from $ep$ reactions
for the same reason. This contrasts with
$p\bar{p}$ and $pp$ reactions, where gluon fragmentation is very important
because a gluon or quark from one (anti)proton can
probe a gluon from the other (anti)proton directly and with a much stronger coupling.

\subsection{Detected hadron mass effect}
\label{ifhadmass}

The production rate of the detected hadron falls as its mass $m_h$ increases
due to the reduction in the size of the available phase space. This effect is particularly
pronounced at small $x_p$ and low $Q$, where $m_h$ cannot be neglected relative to
the hadron's spatial momentum. Treatment of the hadron mass effect in the timelike case was covered in 
Ref.\ \cite{Albino:2005gd}; here we derive the modification to Eq.\ (\ref{xsinfacttheor})
in the spacelike case. The result is essentially 
equivalent to that of Ref.\ \cite{Albino:2005gd} after making the replacement $s \rightarrow Q^2$,
where $\sqrt{s}$ is the c.m.\ energy of the $e^+ e^-$ system.

In general, the scaling variables of the factorization theorem are given
by ratios of the light cone momenta. To find the general relation between 
the true scaling variable of fragmentation
and the measured variable $x_p$ in the presence of hadron mass,
we work in the class of frames in which the spatial momenta of the
virtual photon and the detected hadron are 
parallel, but is otherwise completely general. It contains, but is not limited to,
the Breit frame, which is achieved by a boost in the direction of the two momenta. 
The 3-axis is chosen to be aligned anti-parallel with this 
direction, with no loss of generality. In light cone coordinates
$V=(V^+,V^-,{\bf V}_T)$, where $V^\pm=(1/\sqrt{2})(V^0 \pm V^3)$ and ${\bf V}_T=(V_1,V_2)$, we then have
\be
q=\left(-\frac{Q^2}{2q^-},q^-,{\bf 0}\right),
\ee
and the momentum of the detected hadron with non-zero mass in terms of the scaling variable
$\xi_p=p_h^-/q^-$, which is invariant with respect to boosts along the 3-axis, is 
\be
p_h=\left(\frac{m_h^2}{2\xi_p q^-},\xi_p q^-,{\bf 0}\right).
\ee
This immediately implies that $\xi_p$ is related to the measured variable $x_p$ through
\be
x_p=\xi_p \left(1-\frac{m_h^2 }{Q^2 \xi_p^2}\right).
\ee
In Eq.\ (\ref{xsinfacttheor}), the partonic momentum must be chosen as
\be
p'=\frac{z}{x_p}(0,p^-_h ,{\mathbf 0}),
\ee
and $\xi_p$ must replace $x_p$ everywhere. The left hand side,
$d {\mathcal O} ^{{\rm proton}\ h}/d\xi_p$, is related to 
the experimentally measured quantity $d {\mathcal O}^{{\rm proton}\ h}/ dx_p$ by
\be
\frac{d {\mathcal O}^{{\rm proton}\ h}}{dx_p}(x,x_p,Q^2)=\frac{1}
{1+\frac{m_h^2 }{Q^2 \xi_p^2(x_p)}}\ \frac{d{\mathcal O}^h}{d\xi_p}(x,\xi_p(x_p),Q^2).
\ee
This normalization of the theoretical cross section agrees with one which has already been proposed 
\cite{med}, and applied in analyses of experimental data \cite{Breitweg:1999nt,Dixon:1999vb}, 
up to terms of $O((m_h^2/(\xi_p^2 Q^2))^2)$.

In principle, the effect of the initial state proton mass, which is most important
at large $x$ and low $Q$, should also be accounted for. However,
since the data we will study are mostly extracted at small $x$ values,
and since this effect modifies the numerator and
denominator of Eq.\ (\ref{XSmeas}) in similar ways, we will neglect it.

\section{Comparisons with HERA data}
\label{compdata}

In this section we present our numerical results for the
single hadron inclusive production measurements
from H1 and ZEUS. The kinematic regions of these data
are discussed in the appendix. In FF fits, uncertainties at small $x_p$, such as
higher twist effects, quark and hadron mass effects and
unresummed soft gluon logarithms in the evolution of the FFs, render
the theoretical calculations for hadron production data from $e^+ e^-$
reactions unreliable when the scaled momentum, given in this case by $x_p=2p_h /\sqrt{s}$,
where $p_h$ is the c.m.\ momentum of the detected hadron, falls below 0.1.
Because of the resulting uncertainties in the FFs at small $x_p$, and because
$ep$ reaction data suffer from similar uncertainties at small $x_p$,
we only study $ep$ reaction data for which $x_p>0.1$. Cross sections are calculated to NLO
in the $\overline{\rm MS}$ scheme using
the CYCLOPS software \cite{Graudenz:1996yg}. We set the number of
active quark flavours $n_f=5$. To account for the initial state
proton, we use the CTEQ6M PDF set of Ref.\ \cite{Pumplin:2002vw}
unless otherwise stated. We use their value
$\Lambda_{\rm QCD}^{(5)}=226$ MeV. Although this does not coincide with the
values at which the various FF sets are obtained, within this range of values
the dependence on $\Lambda_{\rm QCD}^{(5)}$ is rather small. 
The factorization / renormalization scale is chosen as $\mu=Q$ unless stated otherwise.
The detected hadron's mass $m_h$
is set to zero unless otherwise stated.

\subsection{Scaled momentum distributions}

\begin{figure}[h!]
\begin{center}
\includegraphics[width=8.5cm]{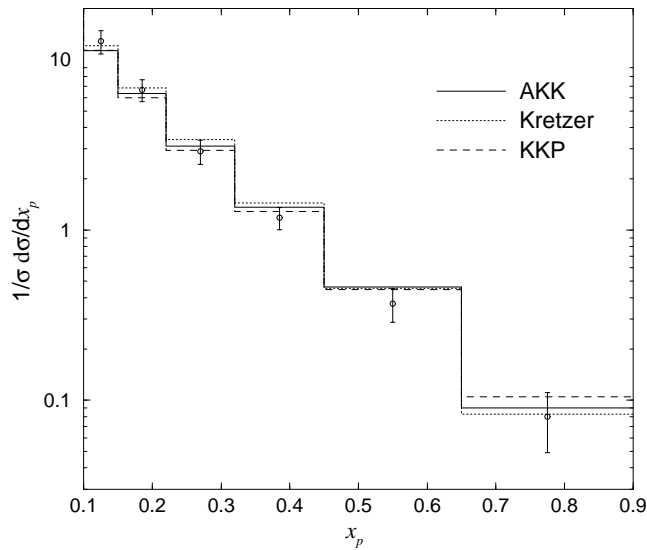}
\caption{Comparisons of theoretical predictions using the AKK, Kretzer
and KKP FF sets with the $x_p$ distribution from ZEUS \cite{Derrick:1995xg}.}
\label{ZEUSold}
\end{center}
\end{figure}

In this subsection we compare theoretical predictions with 
single hadron inclusive production $x_p$ distributions measured
by H1 \cite{Adloff:1997fr} (see Fig.\ \ref{H1oldlimits} for the kinematical constraints) and 
ZEUS \cite{Derrick:1995xg} (see Fig.\ \ref{ZEUSoldlimits}).
The predictions generally agree well with the 
ZEUS data (Fig.\ \ref{ZEUSold}). 
The predictions using the 
Kretzer FF set \cite{Kretzer:2000yf} are similar to those
in Ref.\ \cite{Nadolsky:2000ky}, where the CTEQ5M1 PDF set was used. 
A similar comparison was performed in Ref.\ \cite{Graudenz:1996an} 
using the BKK FF set \cite{Binnewies:1994ju}, and the agreements were good when the 
CTEQ3M and MRSA$^\prime$ PDF sets were used.
\begin{figure}[b!]
\begin{center}
\includegraphics[width=8.5cm]{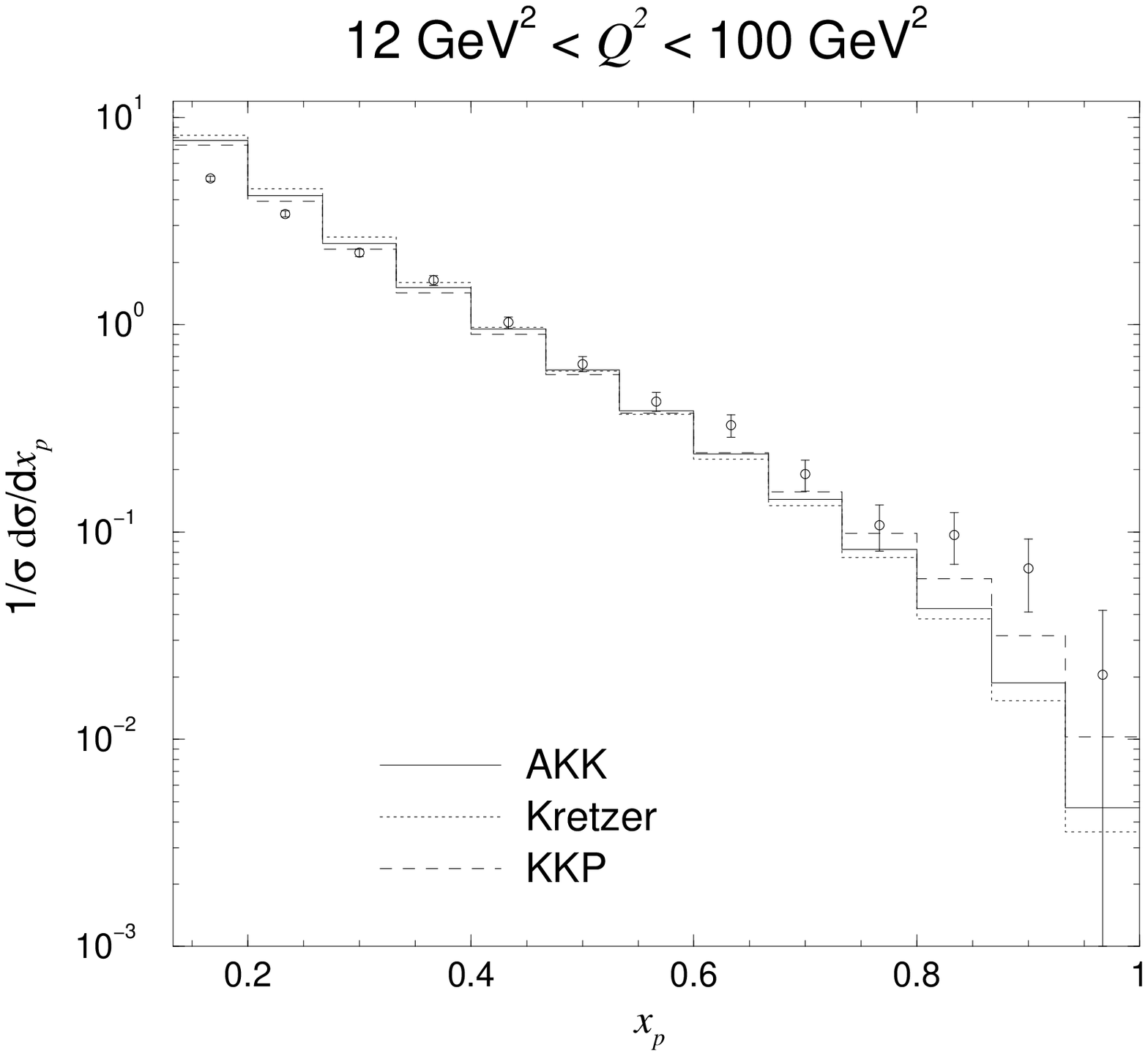}
\caption{As in Fig.\ \ref{ZEUSold}, for the low $Q$ H1 $x_p$ distribution \cite{Adloff:1997fr}.}
\label{H1old_xfL}
\end{center}
\end{figure}
\begin{figure}[b!]
\begin{center}
\includegraphics[width=8.5cm]{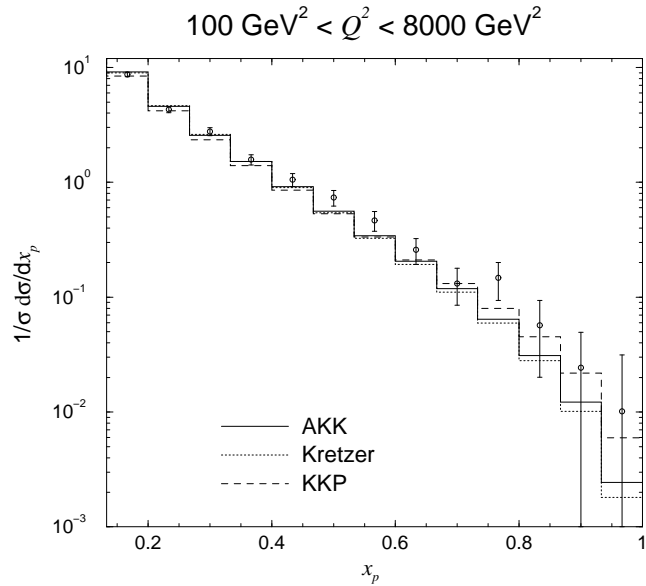}
\caption{As in Fig.\ \ref{ZEUSold}, for the high $Q$ H1 $x_p$ distribution \cite{Adloff:1997fr}.}
\label{H1old_xfH}
\end{center}
\end{figure}
For both the H1 (Figs.\ \ref{H1old_xfL} and \ref{H1old_xfH}) and ZEUS data,
the predictions using the KKP FF set 
are the most gradual in $x_p$, while the Kretzer predictions are the steepest. 
The predictions from the AKK and Kretzer sets are quite similar, particularly at large $x_p$ and
for all $x_p$ values of the high $Q$ H1 data (Fig.\ \ref{H1old_xfH}).
The uncertainty from the freedom in the choice of FF set
is largest at large $x_p$, since the data from $e^+ e^-$ reactions 
is most inaccurate and most scarce at large $x_p$.
The predictions for the low $Q$ H1 data (Fig.\ \ref{H1old_xfL})
show an undershoot
at large $x_p$. This behaviour may result from unresummed logarithms at large $x_p$ in the 
partonic cross section, since resummation tends to enhance the cross section. 
The overshoot from the low $Q$ H1 data at small $x_p$ may be due to the theoretical errors 
in $ep$ reaction data discussed above. Indeed, better agreement is found at small $x_p$ with the
high $Q$ H1 data (Fig.\ \ref{H1old_xfH}),
where resummation is less necessary and where higher twist and mass effects are significantly reduced.

\begin{figure}[h!]
\begin{center}
\includegraphics[width=8.5cm]{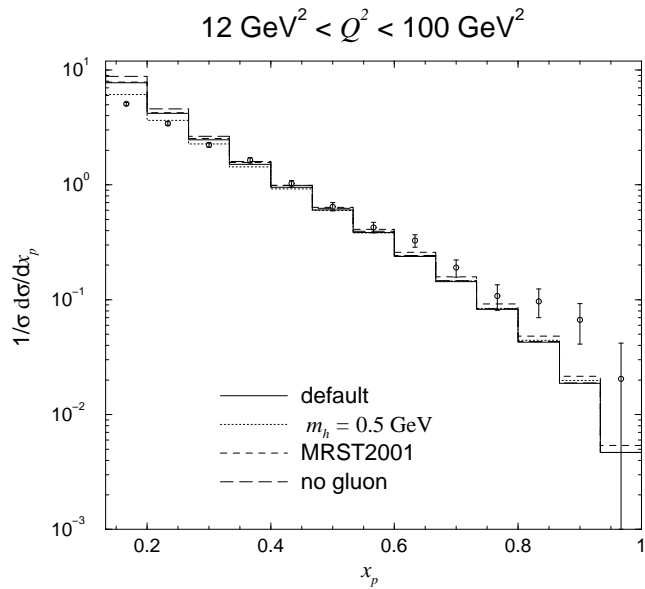}
\caption{As in Fig.\ \ref{H1old_xfL}, using only the AKK FF set.
The modifications to the default predictions (solid line)
arising from the replacement of the CTEQ6M PDF set by the
MRST2001 PDF set of Ref.\ \cite{Martin:2001es}, from the removal
of the evolved gluon, and from the incorporation of the hadron mass effect are shown.}
\label{H1old_combine_xfL}
\end{center}
\end{figure}

\begin{figure}[h!]
\begin{center}
\includegraphics[width=8.5cm]{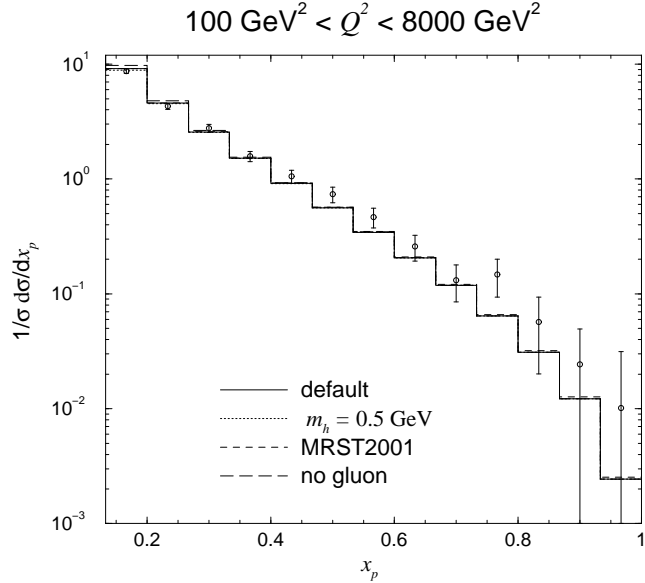}
\caption{As in Fig.\ \ref{H1old_combine_xfL}, for the high $Q$ H1 $x_p$ distribution.}
\label{H1old_combine_xfH}
\end{center}
\end{figure}

We now study various modifications to the predictions for the low and high $Q$ H1 data in order to 
understand the effect of increasing $Q$ on the theoretical and propagated experimental errors.
First we modify our theoretical approach to incorporate the detected hadron mass
according to the method of subsection \ref{ifhadmass}. 
Since the hadron sample is dominated by pions, 
the ``average'' hadron mass is expected to be around
$m_h=0.2-0.3$ GeV. However, to exaggerate the effect 
of hadron mass for illustration, we choose the larger 
value $m_h=0.5$ GeV.
At small $x_p$, this effect improves the description of the 
low $Q$ H1 data (Fig.\ \ref{H1old_combine_xfL}), 
while making negligible difference to the high $Q$ H1 data (Fig.\ \ref{H1old_combine_xfH})
over the whole $x_p$ range. However, this improvement should not be taken too
seriously, since other low $Q$, small $x_p$ effects may also be relevant. In addition,
the FFs from the various sets are artificially suppressed at small $x_p$ since the hadron mass effect
was not accounted for in the analyses of Refs.\ \cite{Kniehl:2000fe,Kretzer:2000yf,Albino:2005me}. 

The error due to the freedom in the choice of PDF set, which we determine by 
calculating the predictions using the MRST2001 PDF set
\cite{Martin:2001es}, is rather small, particularly for intermediate $x_p$ values and for
the high $Q$ data.

\begin{figure}[h!]
\begin{center}
\includegraphics[width=8.5cm]{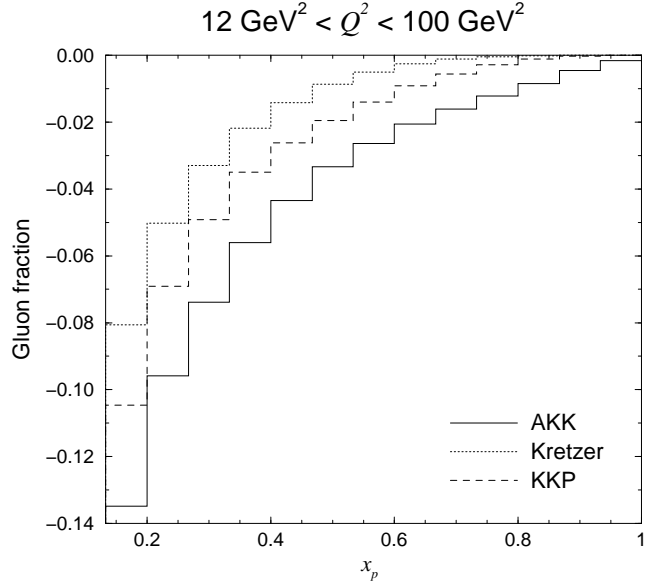}
\caption{Ratios of the evolved gluon contribution to the H1 low $Q$ measurement 
to the full measurement, calculated using the AKK, Kretzer and KKP FF sets.}
\label{H1old_gluon_xfL}
\end{center}
\end{figure}

The gluon contribution (also shown in Fig.\ \ref{H1old_combine_xfL})
is clearly negative, although the evolved gluon FF is positive. This quantity is calculated by
setting the evolved quark FFs to zero in $F^{{\rm proton}\ h}$. Although
it is scheme and scale dependent, its definition is the same for all 3 FF
sets, and therefore its variation with respect to the choice of set is due only to the different choices
of $e^+ e^-$ reaction data used in the fits.
In general, the gluon fragmentation is unimportant, particularly away from the 
smaller $x_p$ range and for the high $Q$ H1 measurements. 
For the low $Q$ H1 data (Fig.\ \ref{H1old_gluon_xfL}), the uncertainty
from the gluon fragmentation from its average is about $\pm$4\% at the
smallest $x_p$ range and about $\pm$2\% at $x_p \approx 0.5$.
This reduces to $\pm$2\% and $\pm$1\% at the same respective $x_p$ values
for the high $Q$ H1 data. The gluon FF is least important for the Kretzer 
predictions, and most important for the AKK ones.

\begin{figure}[h!]
\begin{center}
\includegraphics[width=8.5cm]{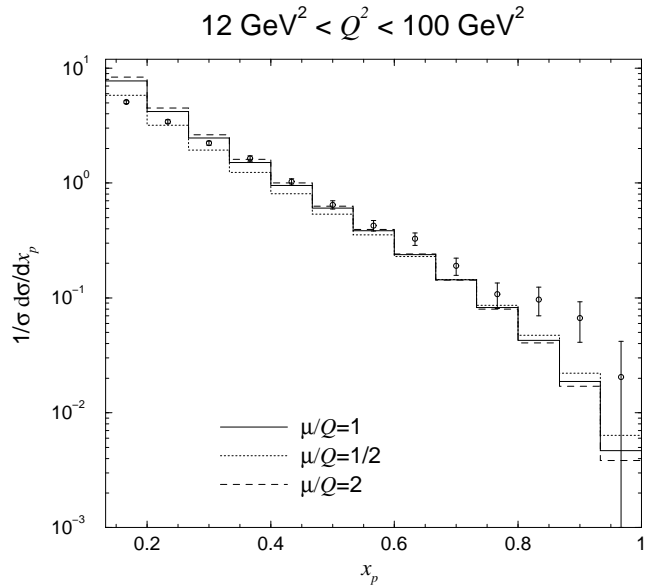}
\caption{As in Fig.\ \ref{H1old_combine_xfL}, for the modifications arising from
scale variation.}
\label{H1old_scale_xfL}
\end{center}
\end{figure}

\begin{figure}[hb!]
\begin{center}
\includegraphics[width=8.5cm]{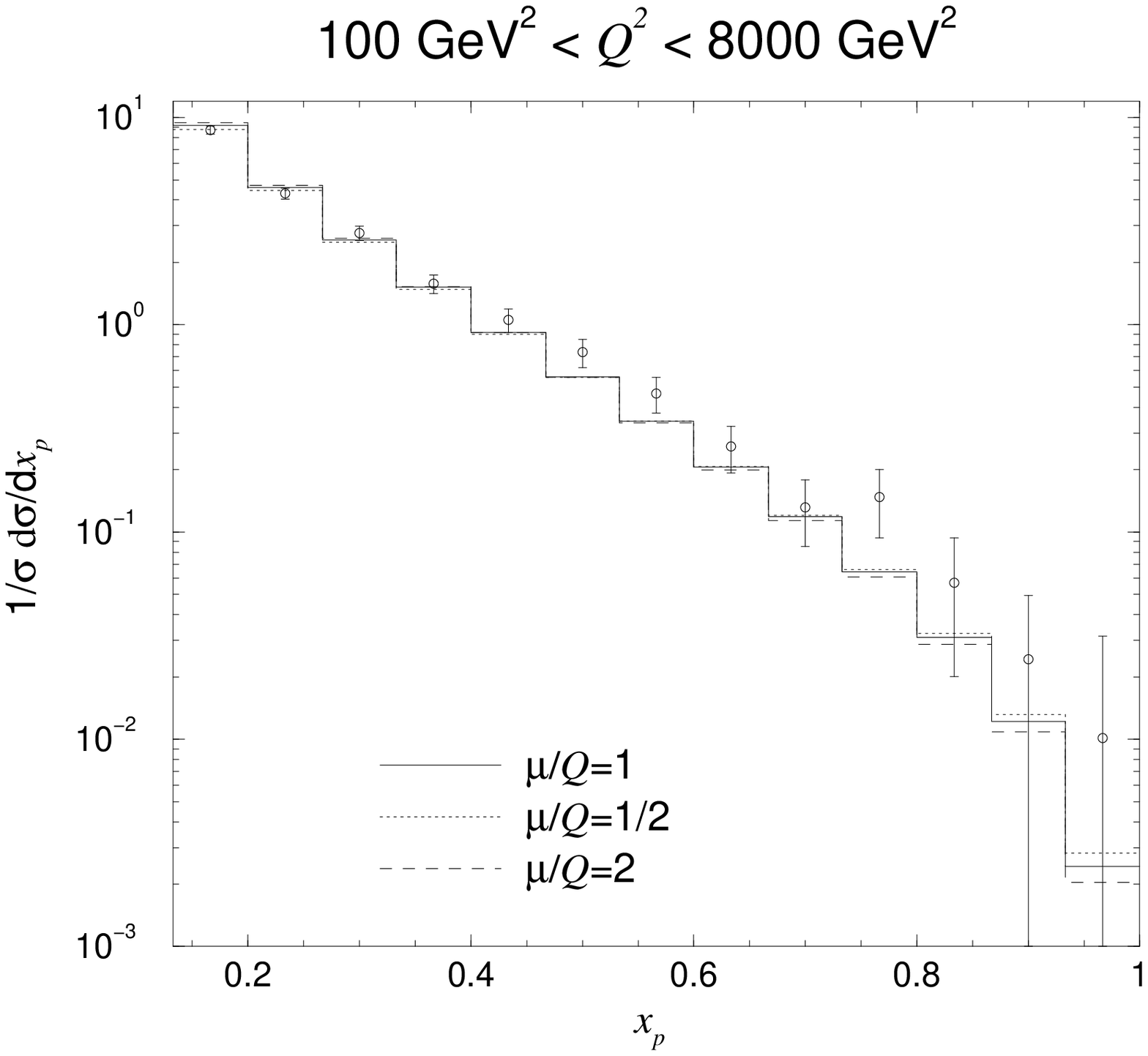}
\caption{As in Fig.\ \ref{H1old_combine_xfH}, for the modifications arising from
scale variation.}
\label{H1old_scale_xfH}
\end{center}
\end{figure}

For the low $Q$ H1 data, 
the uncertainty from the freedom in the scale choice is largest at the smaller and larger 
$x_p$ values (Fig.\ \ref{H1old_scale_xfL}). In addition, since $Q$ is low, our neglect of
charm quark threshold effects is expected to contribute significant errors at small $x_p$.
These uncertainties are most likely dominated by unresummed logarithms at small and large $x_p$
discussed above since, for the high $Q$ H1 data (Fig.\ \ref{H1old_scale_xfH}),
the error at small $x_p$ is much smaller,
while the error at large $x_p$ still remains sizeable. 
In general, increasing the scale
steepens the drop in the cross section with increasing $x_p$.

\begin{figure}[ht!]
\begin{center}
\includegraphics[width=8.5cm]{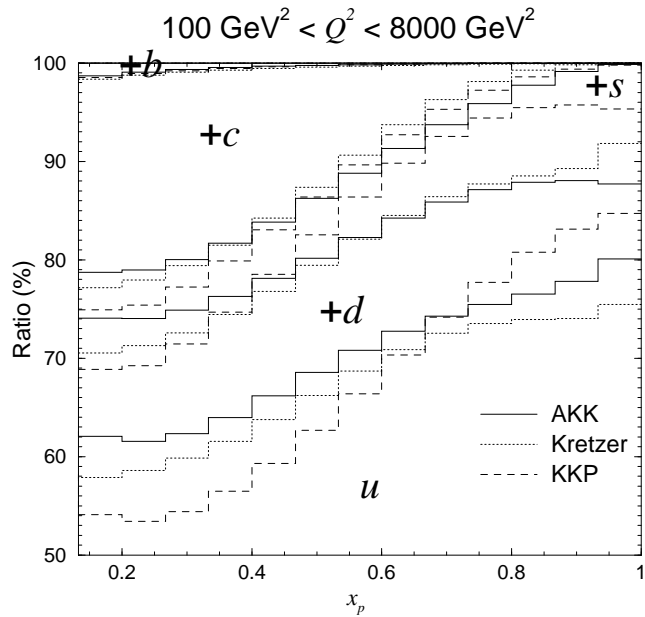}
\caption{The ratios of the quark tagged components of the cross section to the untagged cross section for the high $Q$ H1 data, 
using the AKK, Kretzer and KKP FF sets. The lowest 3 curves show the contribution from the $u$ quark
tagged component only, the next 3 curves above the sum of the $u$ and $d$ components, the next 3 $u$, $d$ 
and $s$ etc.}
\label{H1old_charge_xfH}
\end{center}
\end{figure}

\begin{figure}[h!]
\begin{center}
\includegraphics[width=8.5cm]{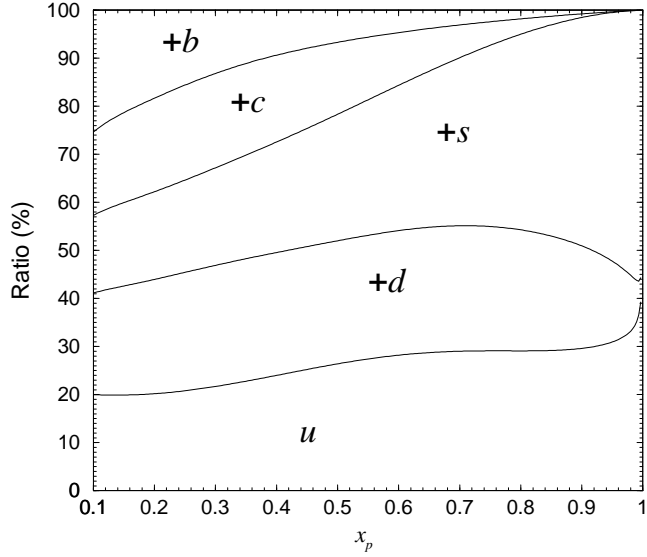}
\caption{The ratios of the quark tagged components of the $e^+ e^- \rightarrow h+X$ cross section, where $h$ is any light charged hadron, 
to the untagged cross section, at $\sqrt{s}=91.2$ GeV and using the AKK FF set.}
\label{qq_sep}
\end{center}
\end{figure}

To determine how the relative importances of the fragmentations from the various quark flavours 
differ between the $ep$ and $e^+ e^-$ reaction data, 
we study the quark flavour tagged components of the cross section in Fig.\ 
\ref{H1old_charge_xfH} (the low $Q$ predictions are not considered since, as we have just seen, 
the theoretical errors are larger), 
and compare with the quark tagged cross sections versus $x_p$
in $e^+ e^-$ reactions (Fig.\ \ref{qq_sep}). The latter reaction was calculated using the method 
of Ref.\ \cite{Albino:2005me}.
As anticipated from the valence structure of the initial state proton in subsection \ref{compepem}, 
the contribution to the overall fragmentation from the 
$u$ quark fragmentation constitutes a significant amount (50\% or more) of the H1 and ZEUS data, 
while the contribution from $d$ quark fragmentation is much less. In the $e^+ e^-$ reaction data,
the $u$ and $d$ quark
fragmentations feature in roughly equal proportions since their
FFs and electroweak charges are similar, and together contribute 50\% or less to the production.
Fragmentation from $s$ quarks is more important in $e^+ e^-$ reactions, particularly at large $x_p$,
while fragmentation from the $c$ quark constitutes similar fractions in  
both reactions. Generally, the part of the
fragmentation arising from the $b$ quark is small due to its small charge and high mass. However, while
it can be relevant in $e^+ e^-$ reactions at small $x_p$, it is always negligible in $ep$ reactions
due to its low density in the proton. 
In the single hadron inclusive production data for $ep$ reactions at large $x_p$,
the contribution to the overall fragmentation from the 
$s$ quark is more important in the predictions of AKK 
and Kretzer than in the KKP predictions. 
On the other hand, as expected, all FF sets lead to similar contributions
from the $u$, $d+s$ and $c$ quark fragmentations, and $b$ quark fragmentation is always
negligible.

\begin{figure}[ht!]
\begin{center}
\includegraphics[width=8.5cm]{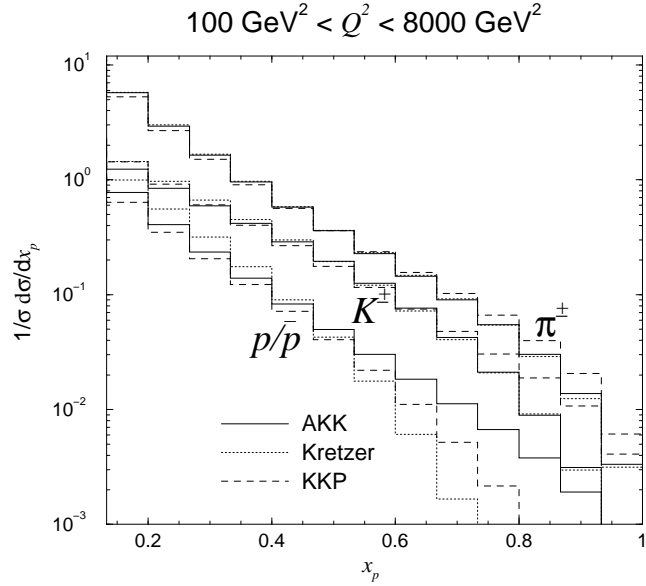}
\caption{The individual
hadron species constituting the sample for the high $Q$ H1 data, using the AKK, Kretzer
and KKP FF sets. Each curve is for a single hadron, not a summation of hadrons.}
\label{H1old_hadsep2_xfH}
\end{center}
\end{figure}

The relative importances of the fragmentations into the various light
charged hadrons in $ep$ reactions can be determined from the composition of the detected
hadron sample with respect to the hadron species (Fig.\ \ref{H1old_hadsep2_xfH}). 
(The Kretzer $p/\bar{p}$ FFs were calculated by subtraction of the $\pi^\pm$ and $K^\pm$ FFs 
from the FFs for all light charged hadrons --- no $p/\bar{p}$ production data was used 
in the extraction of the Kretzer FFs.) The uncertainty in the different yields
is estimated by the spread of the results for the different FF sets, and is largest at large $x_p$
and smallest at intermediate $x_p$. 
The AKK and Kretzer sets give rather similar descriptions of the $\pi^\pm$ and $K^\pm$ yields for
all $x_p$ values shown, while the KKP set gives larger yields at large $x_p$.
Fragmentation to $p/\bar{p}$ at large $x_p$, where all three predictions
differ considerably, is clearly difficult to calculate reliably.
\begin{figure}[ht!]
\begin{center}
\includegraphics[width=8.5cm]{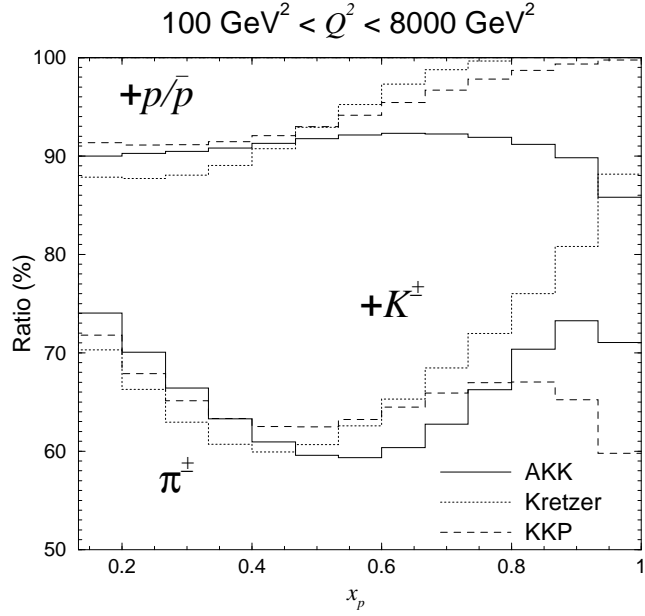}
\caption{The ratios of the individual
hadron species constituting the sample for the high $Q$ H1 data, using the AKK, Kretzer
and KKP FF sets.}
\label{H1old_hadsep2_xfH_ratios}
\end{center}
\end{figure}
\begin{figure}[hb!]
\begin{center}
\includegraphics[width=8.5cm]{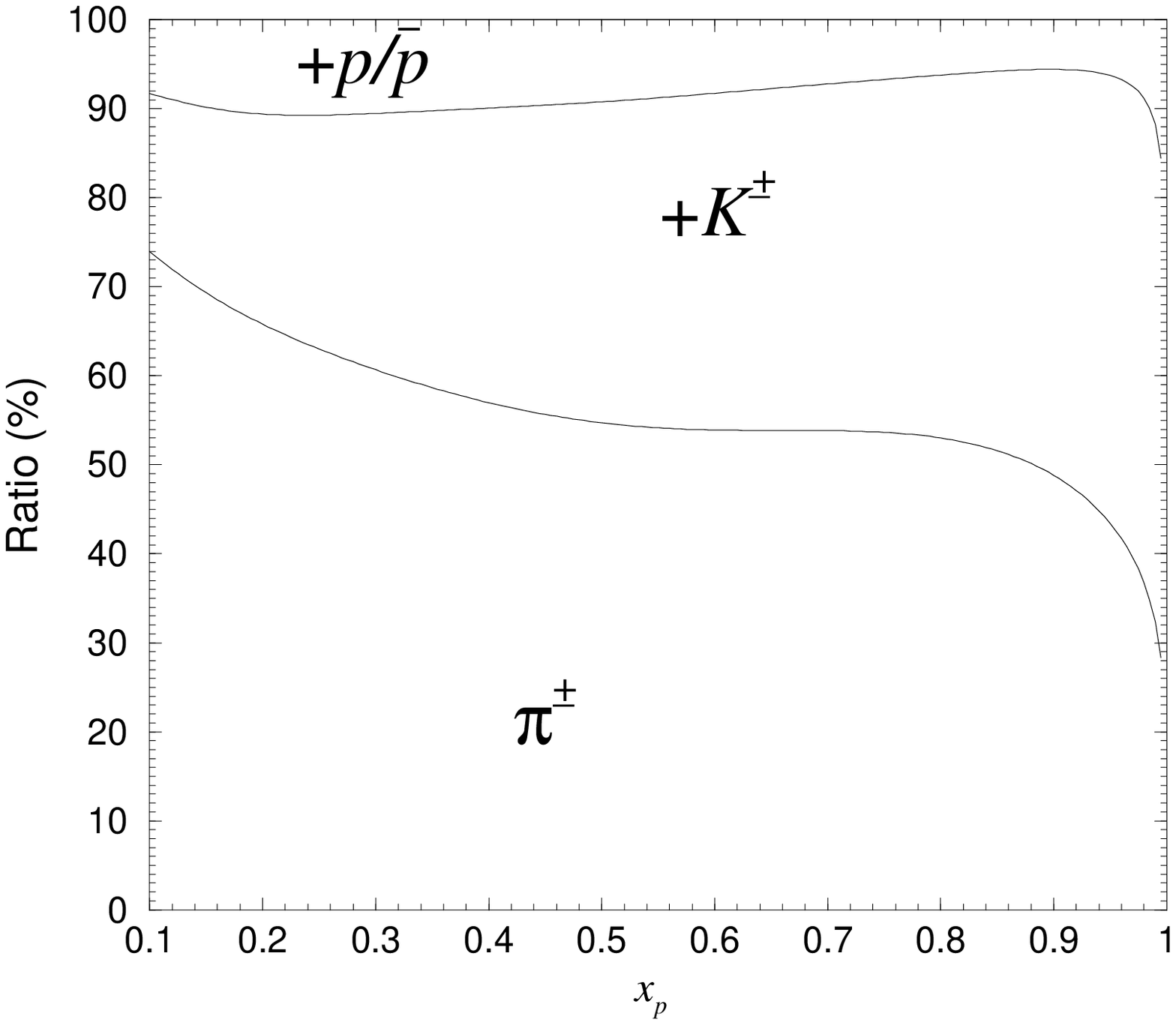}
\caption{The ratios of the individual
hadron species constituting the sample for the $x_p$ distribution 
of the $e^+ e^- \rightarrow h+X$ cross section, where $h$ is any light charged hadron, to the cross section for the full sample, at 
$\sqrt{s}=91.2$ GeV and using the AKK FF set.}
\label{qq_hadsep}
\end{center}
\end{figure} 
The $\pi^\pm$ yield in $ep$ reactions (Fig.\ \ref{H1old_hadsep2_xfH_ratios}), as for $e^+ e^-$ reactions (Fig.\ \ref{qq_hadsep}),
is the largest one due to the low mass of the charged pion.
The fraction of $K^\pm$ is slightly larger
in $e^+ e^-$ reactions than in $ep$ reactions, possibly because
$s$ quark fragmentation is more important in the former data:
The most important source of $K^\pm$ is the $s$ quark, since the other favoured
quark, $u$, has to extract a heavier $s$ quark from the sea.

\subsection{Distributions in photon virtuality}

\begin{figure*}[h!]
\begin{center}
\includegraphics[width=15cm]{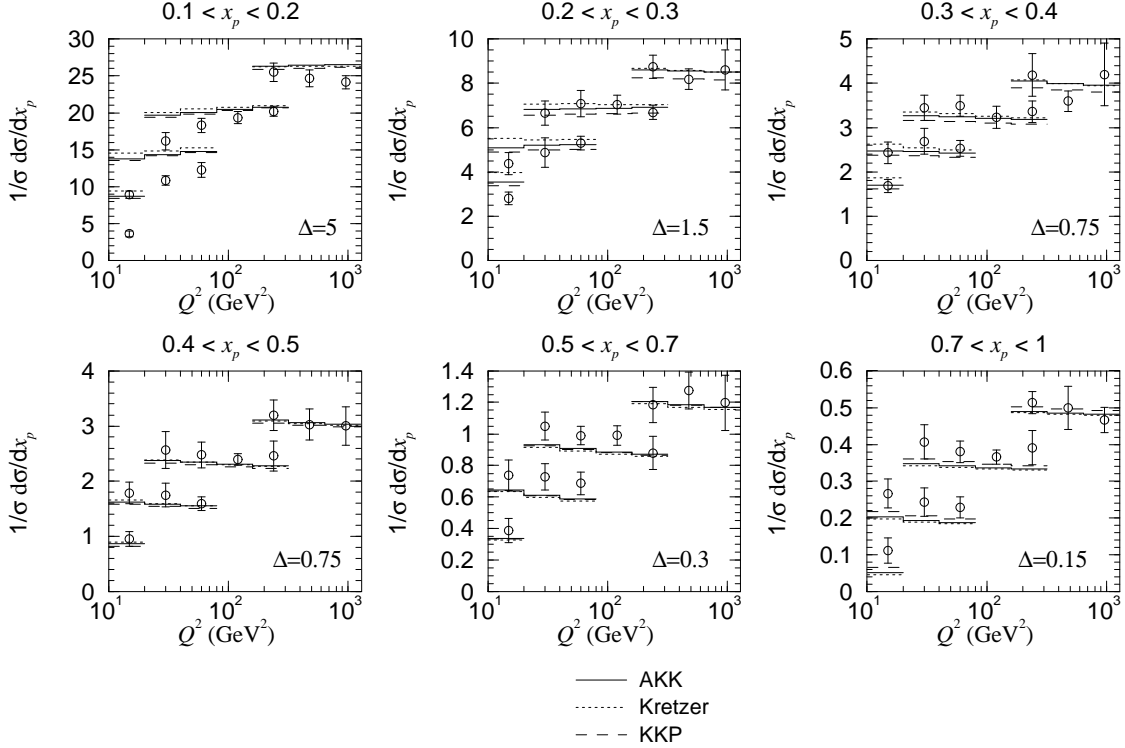}
\caption{Comparisons of theoretical predictions using the AKK, Kretzer
and KKP FF sets with the ZEUS data \cite{Breitweg:1997ra}. 
Each data set is measured in a specific $x$-bin and, together with its predictions, is
shifted upwards relative to the one below by the indicated value for $\Delta$.}
\label{ZEUSold2}
\end{center}
\end{figure*}

\begin{figure*}[h!]
\begin{center}
\includegraphics[width=15cm]{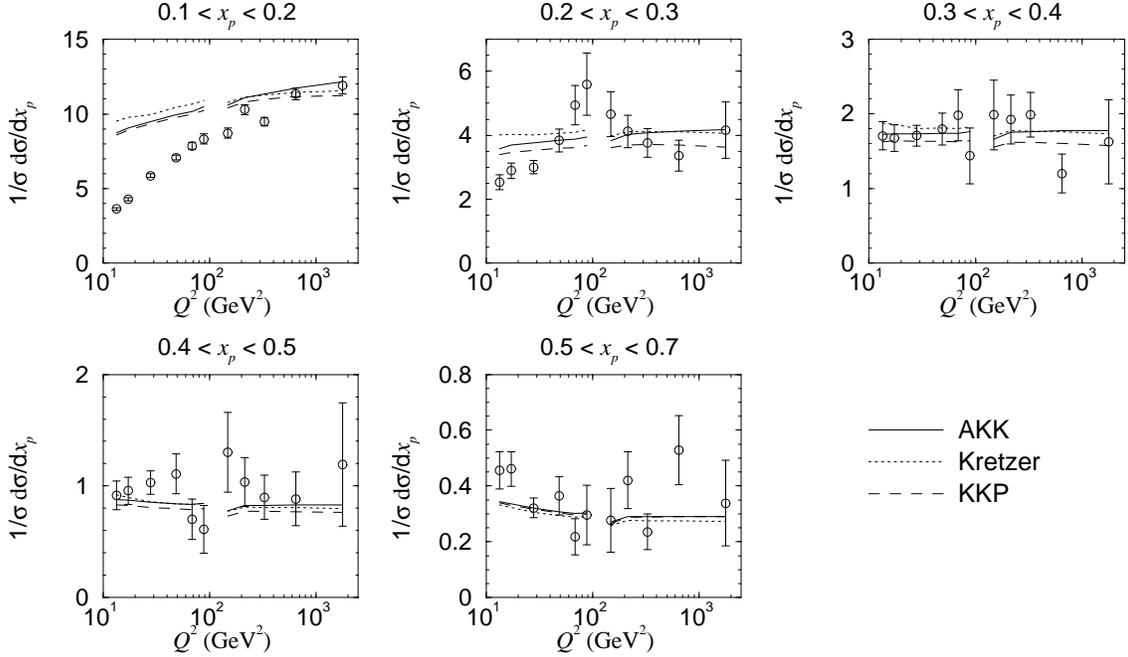}
\caption{As in Fig.\ \ref{ZEUSold2}, for the H1 data \cite{Adloff:1997fr}.}
\label{H1old_Q}
\end{center}
\end{figure*}

Next we compare theoretical predictions with the single hadron inclusive production measurements
at various $Q$ values from H1 \cite{Adloff:1997fr} (see
Fig.\ \ref{H1oldlimits}) and ZEUS
\cite{Breitweg:1997ra} (see Fig.\ \ref{ZEUSold2limits}). The predictions
agree well with the ZEUS data (Fig.\ \ref{ZEUSold2}), except for, at
low $Q$, the overshoot at small $x_p$ and the undershoot at large
$x_p$. Similar behaviour is found with the less precise H1 data (Fig.\ \ref{H1old_Q}).
Note that the theoretical predictions are
rather constant over the whole $Q$ range of both data sets, as foreseen in subsection
\ref{compepem}. Except at the lower $Q$ and smaller $x_p$ region, the AKK predictions tend to 
be closer to the Kretzer predictions than to the KKP ones.

\begin{figure*}[h!]
\begin{center}
\includegraphics[width=10cm]{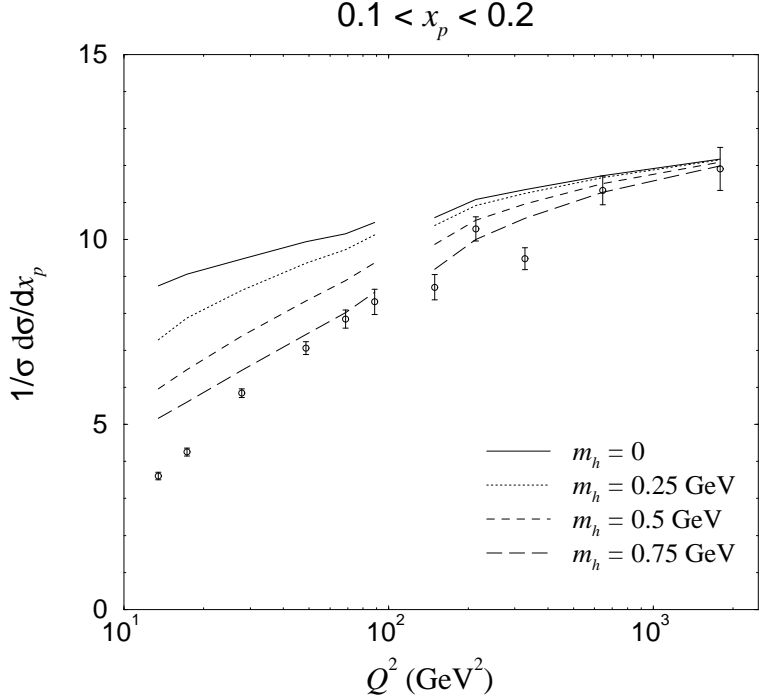}
\caption{As in Fig.\ \ref{H1old_Q} for the data measured in the interval $0.1<x_p <0.2$, 
using only the AKK FF set and for different values of $m_h$.}
\label{H1old_outmass_Q}
\end{center}
\end{figure*}

\begin{figure*}[h!]
\begin{center}
\includegraphics[width=15cm]{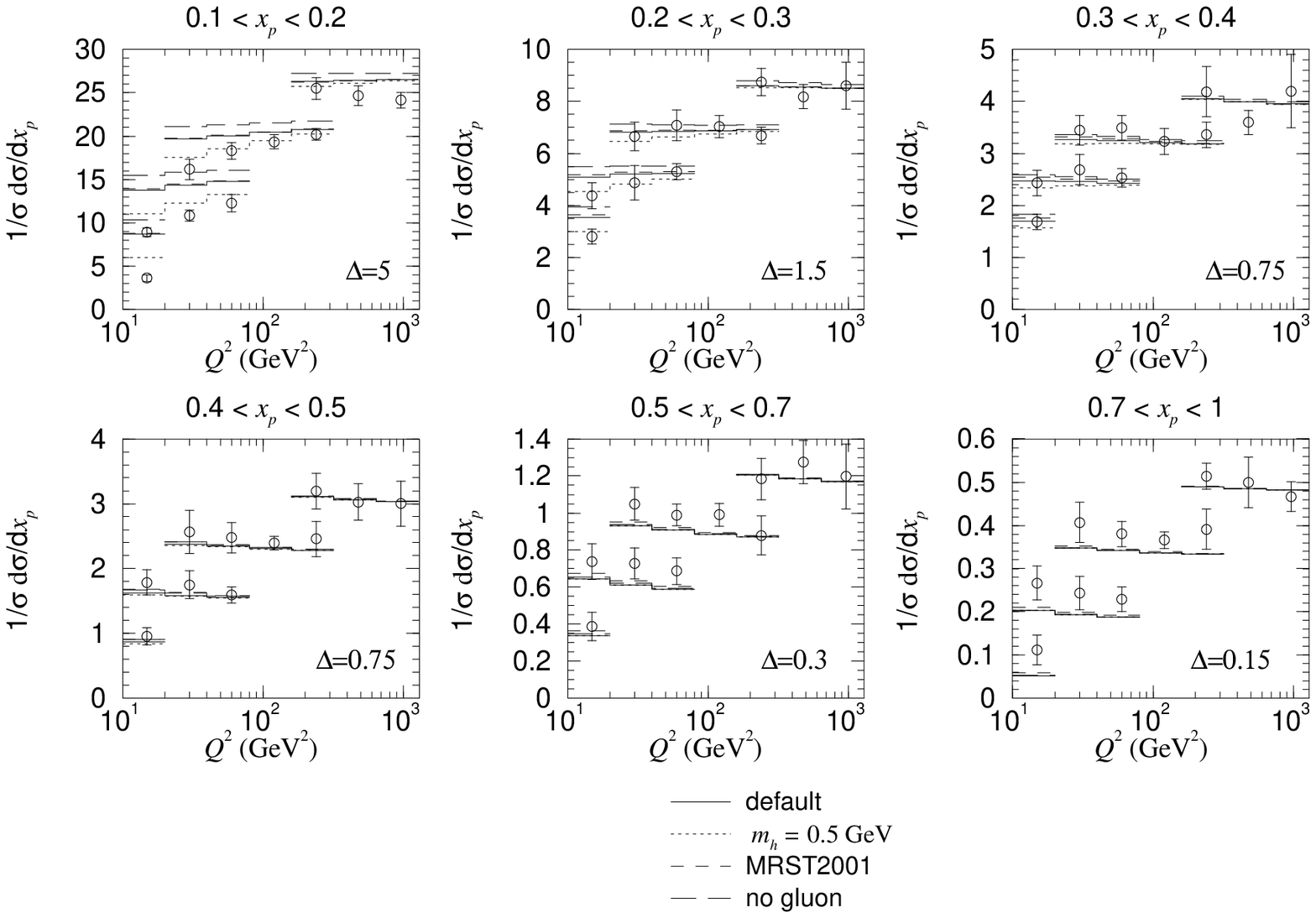}
\caption{As in Fig.\ \ref{ZEUSold2}, using the AKK FF set.
The modifications to the default predictions (solid line)
arising from the replacement of the CTEQ6M PDF set by the
MRST2001 PDF set of Ref.\ \cite{Martin:2001es}, from the removal
of the evolved gluon, and from the incorporation of hadron mass effects are shown.}
\label{ZEUSold2_combine}
\end{center}
\end{figure*}

The hadron mass effect brings the prediction closer to the data (Figs.\ \ref{H1old_outmass_Q}
and \ref{ZEUSold2_combine}) at low $Q$ and small $x_p$. In this region charm threshold effects are expected to be 
important, and this may explain the large average hadron mass required to obtain convergence of the theory with the data.
Good agreement with the H1 data was obtained in Ref.\ \cite{Dixon:1999vb} by essentially choosing 
$m_h=0.66$ GeV. 

\begin{figure*}[h!]
\begin{center}
\includegraphics[width=15cm]{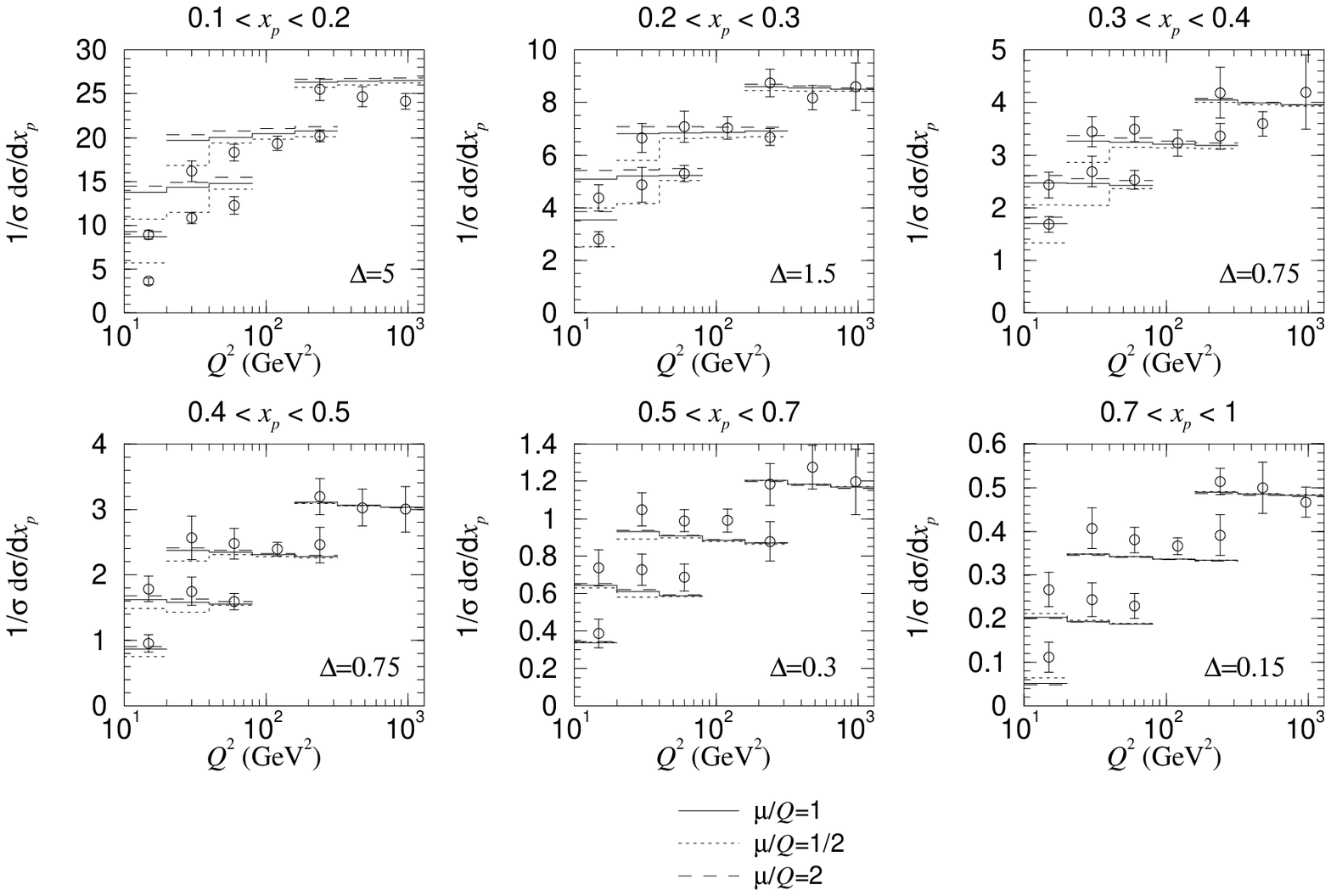}
\caption{As in Fig.\ \ref{ZEUSold2_combine}, for the modifications arising from
scale variation.}
\label{ZEUSold2_scale}
\end{center}
\end{figure*}

The uncertainty from the freedom in the choice of
PDF set for the proton (Fig.\ \ref{ZEUSold2_combine}) is everywhere insignificant. 
At smaller $x_p$ values, the gluon fragmentation 
and the uncertainty with respect to the arbitrary scale choice 
(Fig. \ref{ZEUSold2_scale}) become less relevant with increasing $Q$, 
and are unimportant for all $Q$ at the other $x_p$ values.
The large deviation of the prediction for $\mu=Q/2$ (dotted line)
from the one for $\mu=Q$ is caused by the vanishing of
the $c$ quark FF below threshold. This behaviour is not physical since we have neglected charm mass
effects. The procedure for incorporating these
effects is given in Ref.\ \cite{Nadolsky:2002jr}, which amounts to
retaining the heavy quark mass dependence in the heavy quark flavour   
creation from photon-gluon fusion, and using the same scaling variable
that results in the latter process for the heavy quark flavour excitation.
Furthermore, the matching conditions of Ref.\ \cite{Cacciari:2005ry} must be imposed on the FFs
at the quark flavour thresholds.
In any case, our results at low $Q$ suffer other theoretical errors
mentioned earlier, such as higher twist.

\begin{figure*}[h!]
\begin{center}
\includegraphics[width=15cm]{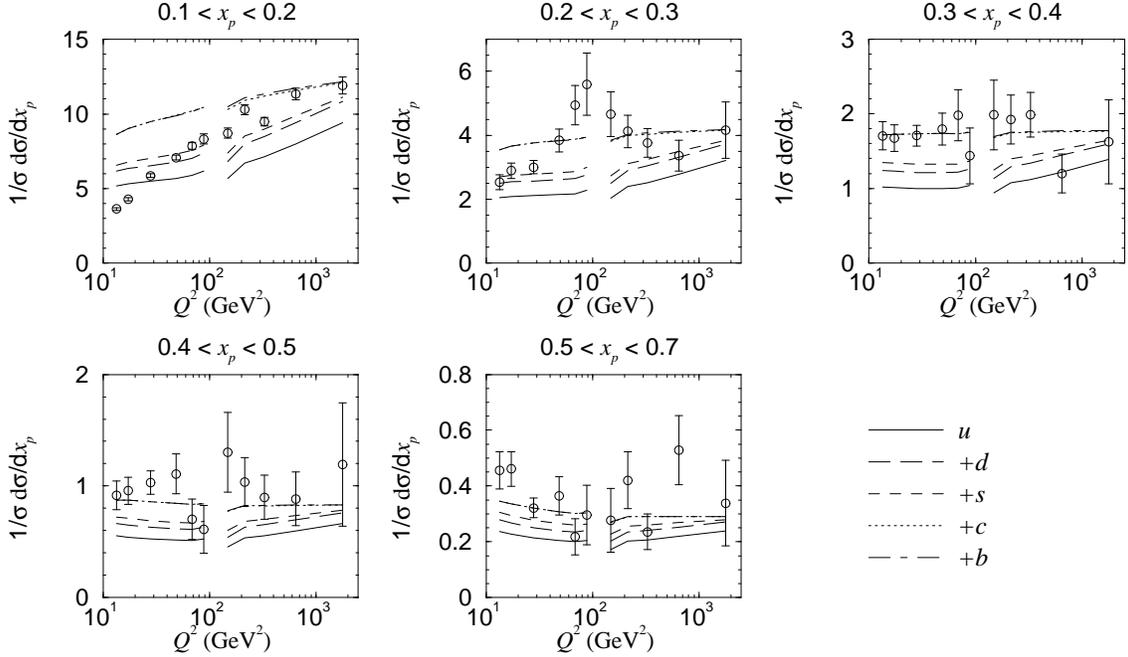}
\caption{As in Fig.\ \ref{H1old_Q}, for the quark tagged components of the cross section using
the AKK FF set.}
\label{H1old_charge_Q}
\end{center}
\end{figure*}

At the lower $Q$ values, the second most important source of fragmentation after the $u$ quark 
is the fragmentation from the $c$ quark (Fig.\ \ref{H1old_charge_Q}),
although, for the H1 data, this falls with rising $Q$
until the $d$ quark fragmentation becomes more important.

\begin{figure*}[h!]
\begin{center}
\includegraphics[width=15cm]{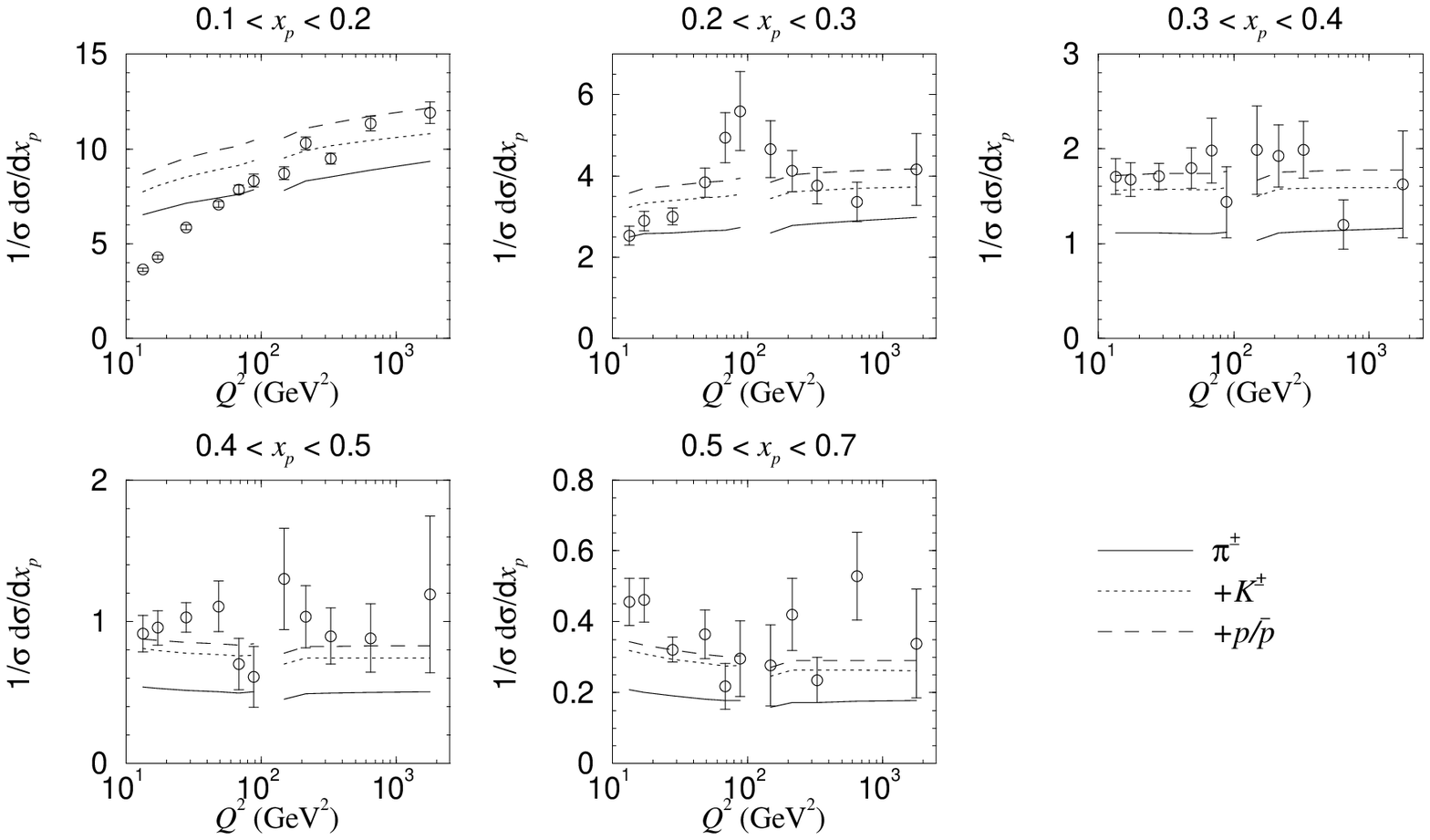}
\caption{As in Fig.\ \ref{H1old_Q}, for the individual
hadron species constituting the sample using the AKK FF set.}
\label{H1old_hadsep_Q}
\end{center}
\end{figure*}

The relative yield of each hadron species does not change significantly
with $Q$ (Fig.\ \ref{H1old_hadsep_Q}), which is expected from perturbation theory at high $Q$.

\newpage
\newpage

\section{Conclusions}
\label{conc}

We have performed a comprehensive analysis of single hadron inclusive
production data at HERA, by calculating the theoretical predictions
using FF sets that were obtained by fitting to accurate $e^+
e^-$ data. In general, good agreement was found using the AKK, Kretzer
and KKP FF sets. However, at low $Q$
and small $x_p$ the predictions overshoot the data, a problem which is
partially remedied by including the detected hadron mass
effect. Unresummed soft gluon logarithms may also contribute to this
discrepancy, as suggested by the increasing variation with respect to the scale for
decreasing $x_p$ and $Q$, as well as higher twist and quark mass effects. 
A more complete treatment which takes into account all these effects is
needed to improve the understanding of fragmentation in this region. At large $x_p$, an
undershoot occurs in the H1 data, which may be avoidable by resumming the logarithms
at large $x_p$, as suggested by the small rise in the variation with respect to the scale
as $x_p$ increases and $Q$ decreases. 
As is the case for $e^+ e^-$ reactions,
gluon fragmentation is not important in $ep$ reaction data, at least for sufficiently large
$x_p$ and $Q$.

Fragmentation from the $u$ quark gives the largest contribution to the overall 
fragmentation in $ep$ reactions, followed by 
$c$ and $d$ quark fragmentation ($c$ quark fragmentation being more important at lower $Q$), 
and finally from the $s$ quark. Fragmentation from the $b$ quark is negligible.
This should be contrasted with the situation in $e^+ e^-$ reactions, where the $u$ and
$d$ quark fragmentations are of similar importance, while fragmentation from $c$
is less important than from $s$, and $b$ quark fragmentation has some relevance at smaller $x_p$.
Therefore, even sufficiently accurate
{\it untagged} $e^+ e^-$ and $ep$ reaction data taken over a large enough range of the kinematic variables
would improve the constraints on the individual quark flavour FFs, although
quark tagging is more valuable for this purpose.

The fractional yields of each of the
light charged hadron species in $ep$ reactions depend strongly on $x_p$, but to a much lesser
extent on $Q$. They are similar to the fractional yields in $e^+ e^-$ reactions.

Relative to the experimental accuracy of the data
sets, the AKK and Kretzer predictions, as well as their quark tagged components and $\pi^\pm$ and
$K^\pm$ yields (but not the evolved gluon fractions and the 
$p/\bar{p}$ yields) are very similar for all data considered. 
At the time of writing, the H1 and ZEUS collaborations are planning an extraction of very accurate
data using, respectively, improved triggering and higher luminosity, which could 
allow for a comparison of the reliablility of the FF sets.

\appendix*
\section{Experimental Cuts}

In this section we present the regions in $(x,Q^2)$ used by the H1 and ZEUS collaborations
from which the measured cross sections are extracted. 
These regions are bounded according to cuts on $x$, $Q^2$,
the squared c.m.\ energy of the virtual photon-proton system,
\be
W^2=(P+q)^2=Q^2 \left(\frac{1}{x}-1\right),
\ee
and the fraction of the energy of the initial electron (we do not distinguish
between electrons and positrons) which is lost in the rest frame of the proton,
\be 
y=\frac{P\cdot q}{P\cdot k}=\frac{Q^2}{xs}.
\ee
A lower bound on the scattered electron's energy
\be
E'=E-Q^2\left(\frac{E}{xs}-\frac{1}{4E}\right),
\label{EprimeEQ2xs}
\ee
where $E$ is the energy of the initial electron,
is sometimes imposed to prevent the scattered electron being falsely identified 
with isolated low energy deposits in the calorimeter while
the true scattered electron passes undetected down the beam pipe.
The H1 collaboration imposes additional cuts \cite{Kant:1995sc}
on the angle of deflection of the electron and struck parton, respectively $\theta_e$ and 
$\theta_p$, to maintain good detector acceptance. In the laboratory frame, these are given in terms
of $x$ and $Q^2$ by
\be
\cos \theta_e
=\frac{xs\left(4E^2-Q^2\right)-4E^2 Q^2}{xs\left(4E^2+Q^2\right)-4E^2 Q^2}
\ee
and
\be
\cos \theta_p =\frac{xs(xs-Q^2)-4E^2 Q^2}{xs(xs-Q^2)+4E^2 Q^2}.
\ee

\begin{figure*}
\begin{center}
\includegraphics[width=15cm]{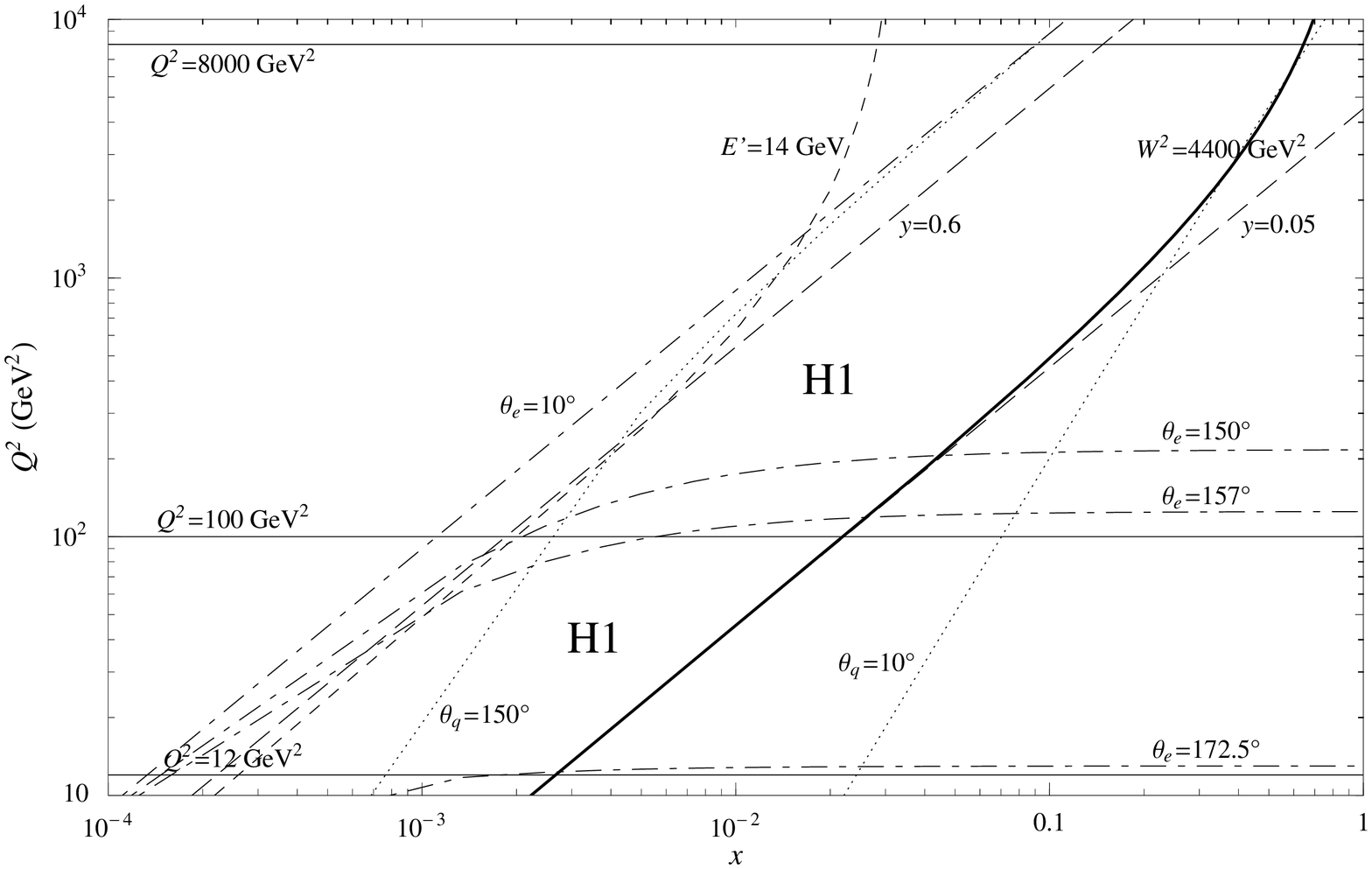}
\caption{Cuts in the $(x,Q^2)$ plane used in the H1 analysis of Ref.\ \cite{Adloff:1997fr}.
The low and high $Q$ regions are each indicated by the label ``H1'',
being bound in each case by the nearest cut to this label. $E=27.5$ GeV
and $\sqrt{s}=300.3$ GeV.}
\label{H1oldlimits}
\end{center}
\end{figure*}

\begin{figure*}
\begin{center}
\includegraphics[width=15cm]{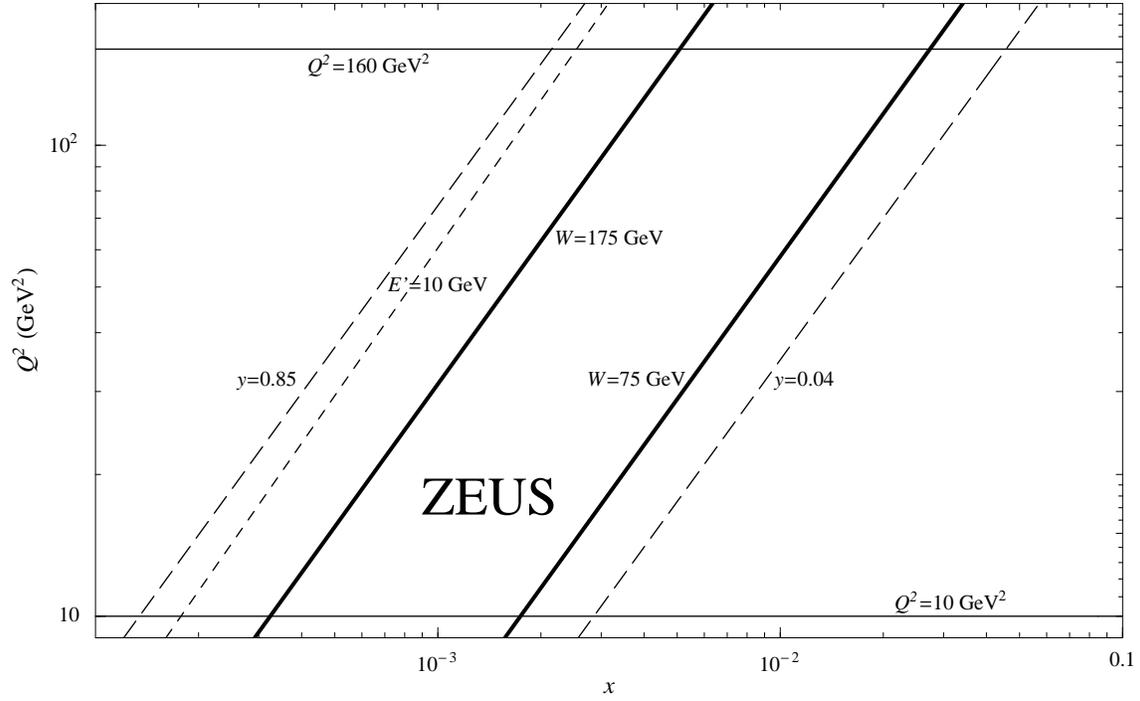}
\caption{Cuts in the $(x,Q^2)$ plane used in the ZEUS analysis of Ref.\ \cite{Derrick:1995xg}.
The measured region is indicated by the label ``ZEUS''. $E=26.7$ GeV and
$\sqrt{s}=296$ GeV.}
\label{ZEUSoldlimits}
\end{center}
\end{figure*}

\begin{figure*}
\begin{center}
\includegraphics[width=15cm]{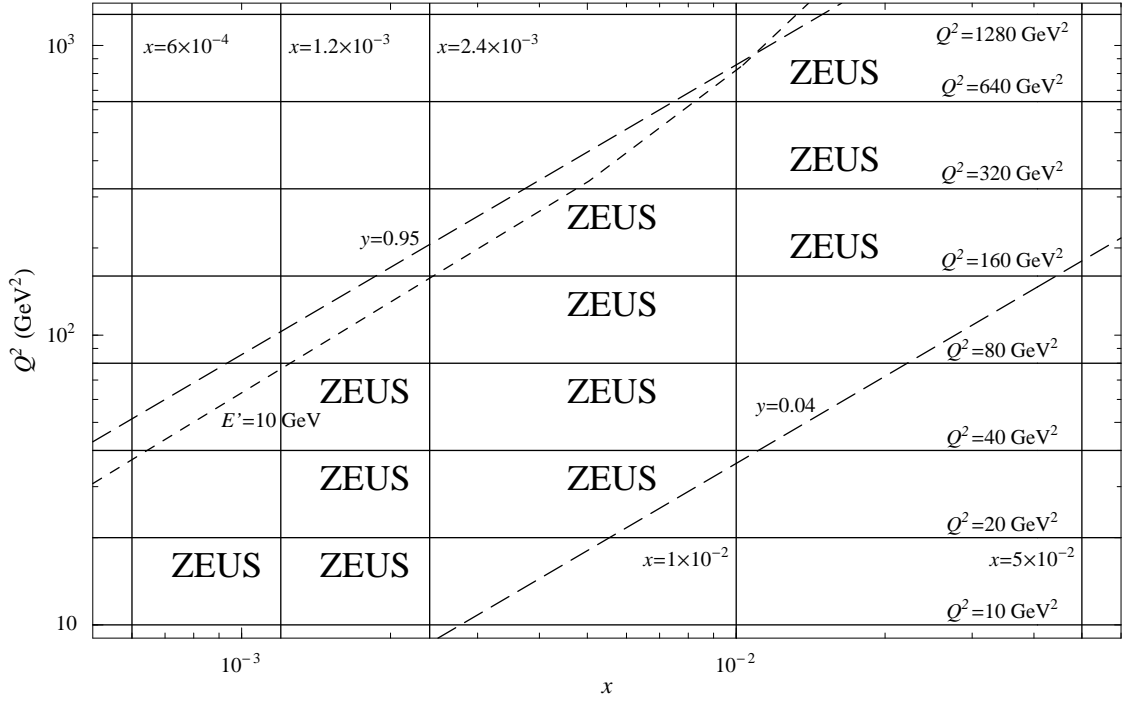}
\caption{Cuts in the $(x,Q^2)$ plane used in the ZEUS analysis of Ref.\ \cite{Breitweg:1997ra}.
The measured regions are indicated by the labels ``ZEUS''. $E=27.5$ GeV and
$\sqrt{s}=300.3$ GeV.}
\label{ZEUSold2limits}
\end{center}
\end{figure*}

The cuts used by H1 in Ref.\ \cite{Adloff:1997fr} are shown in Fig.\ \ref{H1oldlimits}.
The lower bound on $\theta_e=10^\circ$ does not bound any of the regions of measurement.
The $Q$ bins in this analysis are very narrow and are not shown.
The cuts used by ZEUS in Ref.\ \cite{Derrick:1995xg} are shown in Fig.\ \ref{ZEUSoldlimits}.
It is clear that only the cuts in $W$ and $Q$ border the region of measurement.
Figure \ref{ZEUSold2limits} shows that the upper bound on $y$ is irrelevant in the extraction
of the data of Ref.\ \cite{Breitweg:1997ra}.


\end{document}